\documentclass{kluwer}    

\newdisplay{guess}{Conjecture}

\newcommand{\be}{\begin{equation}}
\newcommand{\ee}{\end{equation}}

\begin{document}                                                                                   
\begin{article}
\begin{opening}         
\title{\bf Gravitation and the Local Symmetry Group \\ of Space-Time}

\author{M. \surname{Cal\c cada}\footnote{Instituto de F\'{\i}sica Te\'orica,
Universidade Estadual Paulista, S\~ao Paulo, Brazil} and J. G.
\surname{Pereira}$^\ddagger$\footnote{To whom correspondence should be addressed at Instituto de
F\'{\i}sica Te\'orica, Universidade Estadual Paulista, Rua Pamplona 145,
01405-900\, S\~ao Paulo, Brazil; e-mail: jpereira@ift.unesp.br}}
\runningauthor{Cal\c cada and Pereira}
\runningtitle{Gravitation and the Local Symmetry Group of Space-Time}

\begin{abstract}

According to general relativity, the interaction of a matter field with gravitation requires the
simultaneous introduction of a tetrad field, which is a field related to translations, and a
spin connection, which is a field assuming values in the Lie algebra of the Lorentz group. These
two fields, however, are not independent. By analyzing the constraint between them, it is
concluded that the relevant {\em local} symmetry group behind general relativity is provided by
the Lorentz group. Furthermore, it is shown that the minimal coupling prescription obtained from
the Lorentz covariant derivative coincides exactly with the usual coupling prescription of general
relativity. Instead of the tetrad, therefore, the spin connection is to be considered as the
fundamental field representing gravitation.

\end{abstract}


\end{opening}           

\section{Introduction}

The group of motions of Minkowski spacetime is the ten--parameter Poincar\'e group, the
semi--direct product of the translation and the Lorentz groups. Denoting by
$\{x^a\}$ ($a$, $b$, $c,\dots=1,2,3,4$) the cartesian coordinates of Minkowski space, and by
\be
\eta_{a b} = {\rm diag} (1, -1, -1, -1)
\ee
its metric tensor, an infinitesimal translation of the spacetime coordinates is defined as
\be
\delta_t x^a = - i \epsilon^c \, P_c \, x^a,
\label{trans}
\ee
where $\epsilon^c$ are the translation parameters, and
\be
P_c = - i \frac{\partial}{\partial x^c} \equiv - i \partial_c
\label{transgen}
\ee
are the translation generators. By using these generators, the transformation (\ref{trans}) can
be rewritten in the form
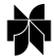
\be
\delta_t x^a = - \epsilon^a.
\label{trans1}
\ee

On the other hand, an infinitesimal Lorentz transformation is defined as
\be
\delta_L x^a = - \frac{i}{2} \, \epsilon^{c d} \, L_{c d} \, x^a,
\label{lore}
\ee
where $\epsilon^{c d}=-\epsilon^{d c}$ are the Lorentz parameters, and
\be
L_{c d} = i (x_c \partial_d - x_d \partial_c)
\label{loregen}
\ee
are the Lorentz generators. By using these generators, the transformation (\ref{lore}) can be
rewritten in the form
\be
\delta_L x^a = - \epsilon^{a}{}_{d} \, x^d.
\label{lore1}
\ee

An interesting property of the Lorentz transformation (\ref{lore}) is that it can be
rewritten formally as a translation~\cite{kibble}. In fact, by using the explicit form of
$L_{c d}$, it becomes
\be
\delta_L x^a = - i \, \xi^c \, P_c \, x^a,
\label{relore}
\ee
which is a translation with
\be
\xi^c = \epsilon^c{}_d \, x^d
\label{constr}
\ee
as the translation parameters. In other words, an infinitesimal Lorentz transformation of the
spacetime coordinates is equivalent to a translation with $\xi^c \equiv \delta_L x^c$ as the
parameter. Actually, this is a property of the Lorentz generators $L_{a b}$, whose action can
always be reinterpreted as a translation. The reason for such equivalence is that, because the
Minkowski spacetime is transitive under translations, every two points related by a Lorentz
transformation can also be related by a translation. Notice that the reverse is not true.

\section{Conserved Quantities}

The con\-ser\-vation laws of energy--momentum and angular--mo\-men\-tum in
special relativity are connected with the Poinca\-r\'e group, the isometry group of Minkowski
spacetime~\cite{trautman}. In fact, according to Noether's theorem~\cite{kopo}, the invariance of
a physical system under a spacetime translation leads to the conservation of the {\em canonical}
energy--momentum tensor, whereas the invariance under a Lorentz transformation leads to the
conservation of the {\em canonical} angular--momentum tensor. When passing to general relativity,
these two tensors are modified by the presence of gravitation. Furthermore, the source of the
gravitational field, the so called {\em dynamical} energy--momentum tensor, turns out to be
a symmetrized version of the modified energy--momentum tensor.

Let us consider the following structure. At each point of spacetime, whose coordinates we denote
by $x^\mu$ ($\mu, \nu, \rho, \dots = 0, 1, 2, 3$), we attach a Minkowski tangent space where both
the Lorentz and the translation groups act locally. It should be remarked that the action of
these two groups are not defined in a curved riemannian spacetime~\cite{wald}. Now according to
the gauge approach to gravitation~\cite{hehl}, the gauge field related to translations shows up
as the non--trivial part of the tetrad field~\cite{kibble}. Denoting by
\be
B = B^a{}_\mu \, P_a \, dx^\mu
\label{bdef}
\ee
the translational gauge potential, which is a connection assuming values in the Lie algebra of
the translation group, the tetrad field is written as~\cite{trans}
\be
h^a{}_\mu = \partial_\mu x^a + c^{-2} B^a{}_\mu,
\label{tetrada}
\ee
where the velocity of light $c$ was introduced for dimensional reasons. Its inverse, denoted by
$h^\rho{}_c$, is defined by the relations
\[
h^a{}_\mu \, h^\mu{}_c = \delta^a{}_b \quad {\rm and} \quad
h^\mu{}_c \, h^c{}_\rho = \delta^\mu{}_\rho,
\]
and is given by an infinite series:
\be
h^\rho{}_c = \partial_c x^\rho - c^{-2} B^\rho{}_c + \dots \; .
\label{itetra}
\ee

On the other hand, the gauge field related to Lorentz transformations is the so called spin
connection $A^{a}{}_{b \mu}$, a connection assuming values in the Lie algebra of the Lorentz
group. Its explicit form is~\cite{dirac}
\be
A^{a}{}_{b \mu} = h^a{}_\rho \left( \partial_\mu h^{\rho}{}_b +
\Gamma^\rho{}_{\nu \mu} \, h^{\nu}{}_b \right) \equiv
 h^a{}_\rho \nabla_\mu h^{\rho}{}_b,
\label{spincon}
\ee
where $\Gamma^\rho{}_{\nu \mu}$ is the Levi--Civita connection of the spacetime metric $g_{\mu
\nu}$, with $\nabla_\mu$ the corresponding covariant derivative. The spacetime and the tangent
space metrics are related by
\be
g_{\mu \nu} = h^a{}_\mu \, h^b{}_\nu \, \eta_{a b}.
\label{ghheta}
\ee

Let us consider now a general matter field $\Psi$ with the action functional
\be
S = \frac{1}{c} \int {\mathcal L} \; d^4x \equiv
\frac{1}{c} \int L \, \sqrt{-g} \ d^4x,
\label{action}
\ee
where $g=\det(g_{\mu \nu})$. According to Noether's theorem, the {\em dynamical}
energy--momentum tensor of the matter field --- that is, the tensor appearing in the right--hand
side of the gravitational field equations --- is given by
\be
{\mathcal T}^\mu{}_a = - \frac{c^2}{h} \, \frac{\delta {\mathcal L}}{\delta B^a{}_\mu},
\label{emt1}
\ee
where $h = \det(h^a{}_\mu) = \sqrt{-g}$. Since the tetrad is linear in the translational gauge
field $B^a{}_\mu$, the functional derivative in relation to $B^a{}_\mu$ can alternatively be
written as a functional derivative in relation to
$h^a{}_\mu$,
\be
{\mathcal T}^\mu{}_a = - \frac{1}{h} \, \frac{\delta {\mathcal L}}{\delta h^a{}_\mu},
\label{emt2}
\ee
which is the form it usually appears in the literature~\cite{weinberg}. On the other hand, the
angular--momentum tensor of the matter field is
\be
{\mathcal J}^\mu{}_{a b} = \frac{1}{h} \, \frac{\delta {\mathcal L}}
{\delta A^{a b}{}_\mu}.
\label{amt}
\ee

It is important to remark that, as the {\em dynamical} energy--momentum tensor (\ref{emt2}) is
automatically symmetric,
\be
h^{a \lambda} \, {\mathcal T}^\mu{}_a =  h^{a \mu} \, {\mathcal T}^\lambda{}_a,
\ee
the {\em total} --- that is, orbital plus spin --- angular momentum tensor is given
by~\cite{weinberg,hayashi}
\be
{\mathcal J}^\mu{}_{a b} = x_a \, {\mathcal T}^\mu{}_b - x_b \, {\mathcal T}^\mu{}_a.
\label{xamt}
\ee
We see in this way that ${\mathcal T}^\mu{}_a$ and ${\mathcal J}^\mu{}_{a b}$ are not
independent tensors. In fact, given the energy--momentum tensor, the expression for the
angular--momentum tensor can immediately be written down.

That ${\mathcal T}^\mu{}_a$ and ${\mathcal J}^\mu{}_{a b}$ are not independent tensors should not
be surprising because the translational gauge potential $B^a{}_\mu$ and the spin connection $A^{a
b}{}_\mu$ are not independent either, as can be seen from Eq.(\ref{spincon}), which gives the
spin connection $A^{a b}{}_\mu$ in terms of the translational gauge potential $B^a{}_\mu$. The
physical reason for this dependency is that both $B^a{}_\mu$ and $A^{a b}{}_\mu$ are produced by
the very same gravitational field.

Let us then look for the inverse relation, that is, let us look for an expression yielding
$B^a{}_\mu$ in terms of $A^{a b}{}_\mu$. By comparing the expressions (\ref{emt1}) and
(\ref{amt}) with (\ref{xamt}), we find immediately that
\be
B^a{}_\mu = c^2 \, A^{a}{}_{b \mu} \, x^b.
\label{bax}
\ee
In fact, from Eq.\ (\ref{amt}), and making use of Eq.\ (\ref{emt1}), we have
\be
{\mathcal J}^\mu{}_{a b} = -
c^{-2} \, {\mathcal T}^\rho{}_c \;
\frac{\delta B^{c}{}_\rho}{\delta A^{a b}{}_\mu}.
\label{amt2}
\ee
But, taking into account that $A^{a b}{}_\mu = -
A^{b a}{}_\mu$, we get from (\ref{bax})
\be
\frac{\delta B^{c}{}_\rho}{\delta A^{a b}{}_\mu} =
c^2 \, \delta^\mu{}_\rho (\delta^c{}_a \, x_b - \delta^c{}_b \, x_a).
\ee
Substituting in (\ref{amt2}), we get exactly the expression (\ref{xamt}) for the
angular momentum ${\mathcal J}^\mu{}_{a b}$.

\section{Minimal Coupling Prescription}

When considering coordinate transformations, only the generators $P_a$ and $L_{a b}$ must be
taken into account. However, in the study of the coupling of a general matter field to
gravitation, other representations of the Lorentz generators show up. For example, under a {\em
local} tangent--space Lorentz transformation, a general matter field
$\Psi(x^\mu)$ will change according to~\cite{ramond}
\be
\delta \Psi \equiv \Psi^\prime(x) - \Psi(x) =
- \frac{i}{2} \, \epsilon^{a b} J_{a b} \Psi,
\label{lt1}
\ee
where $J_{a b}$ is an appropriate generator of the infinitesimal Lorentz transformations. The
most general form of $J_{a b}$ is
\be
J_{a b} = L_{a b} + S_{a b},
\label{fullrep}
\ee
where $L_{a b}$ is the {\em orbital} part of the generator, whose explicit form, given by
(\ref{loregen}), is the same for all fields, and $S_{a b}$ is the {\em spin} part of the
generator, whose explicit form depends on the spin contents of the field $\Psi$. Notice that the
orbital generators $L_{a b}$ are able to act in the spacetime argument of $\Psi(x^\mu)$ due to
the relation
\[
\partial_a = (\partial_a x^\mu) \, \partial_\mu.
\]
By using the explicit form of $L_{a b}$, the Lorentz transformation (\ref{lt1}) can be
rewritten as
\be
\delta \Psi = - \epsilon^{a b} \, x_b \, \partial_a \Psi -
\frac{i}{2} \, \epsilon^{a b}{} S_{a b} \Psi,
\label{lt2}
\ee
or equivalently,
\be
\delta \Psi = - \xi^c \partial_c \Psi -
\frac{i}{2} \, \epsilon^{a b} S_{a b} \Psi,
\label{lt3}
\ee
where use has been made of Eq.\ (\ref{constr}). In other words, the {\em orbital} part of the
transformation can be reduced to a translation, and consequently the Lorentz transformation of
a general field $\Psi$ can be rewritten as a translation plus a strictly spin Lorentz
transformation. It should be remarked that, despite the similarity with a Poincar\'e
transformation, it does not correspond to a transformation of the Poincar\'e group because in
this group the translation and the Lorentz parameters are completely independent. This is
clearly not the case here because of the constraint (\ref{constr}) between the translation
and the Lorentz parameters.

As is well known, the gravitational minimal coupling prescription amounts to replace all
flat--spacetime ordinary derivatives $\partial_a$ by covariant derivatives ${\mathcal D}_a$. The
general definition of covariant derivative is~\cite{livro}
\be
{\mathcal D}_c \Psi = \partial_c \Psi +
\frac{1}{2} \, A^{a b}{}_c \,
\frac{\delta \Psi}{\delta \epsilon^{a b}},
\ee
where $A^{a b}{}_c = A^{a b}{}_\mu \, h^\mu{}_c$. Substituting (\ref{lt2}), we get
\be
{\mathcal D}_c \Psi = \partial_c \Psi -
A^{a b}{}_{c} \, x_b \,  \partial_a \Psi -
\frac{i}{2} \, A^{a b}{}_c \, S_{a b} \, \Psi,
\ee
or equivalently,
\be
{\mathcal D}_c \Psi = (\delta^a{}_c - A^{a}{}_{b c} x^b ) \partial_a \Psi -
\frac{i}{2} \, A^{a b}{}_c \, S_{a b} \, \Psi.
\ee
Then, by making use of Eqs. (\ref{tetrada}) and (\ref{bax}), we can write
\be
{\mathcal D}_c \Psi = h^\mu{}_c \, {\mathcal D}_\mu \Psi,
\ee
with
\be
{\mathcal D}_\mu = \partial_\mu -
\frac{i}{2} \, A^{a b}{}_\mu \, S_{a b}
\ee
the Fock--Ivanenko covariant derivative operator~\cite{fi1,fi2}. Therefore, the minimal
coupling prescription associated with the transformation (\ref{lt2}) can be stated in the form
\be
\partial_c \rightarrow {\mathcal D}_c = h^\mu{}_c \, {\mathcal D}_\mu,
\ee
which is exactly the usual coupling prescription of general relativity. In fact,
as is well known, in the coupling prescription of general relativity the tetrad $h^a{}_\mu$ and
the spin connection $A^{a b}{}_\mu$ are not independent fields. Such a coupling prescription, as
we have shown, can be obtained from a Lorentz covariant derivative with the complete
representation (\ref{fullrep}). In this covariant derivative, the {\em orbital} part of
the Lorentz generators are reduced to a translation, which gives rise to a tetrad that depends on
the spin connection. This reduction, therefore, is the responsible for the constraint between the
tetrad field and the spin connection. The same constraint gives rise also to the relation between
energy--momentum and angular--momentum tensors of a matter field.

\section{Final Remarks}

The basic results of this paper can be summarized in the following way. As is well--known, the
energy--momentum conservation is related to the invariance of the action under a translation of
the spacetime coordinates, and the angular--momentum conservation is related to the invariance of
the action under a Lorentz transformation. However, as the symmetric energy--momentum tensor
${\mathcal T}^\mu{}_a$ and the {\em total} angular--momentum tensor ${\mathcal J}^\mu{}_{a b}$
are not independent quantities, the parameters related to translation and Lorentz transformation
can not be independent either. In fact, they are related by
\be
\xi^a = \epsilon^a{}_b \, x^b,
\label{rela}
\ee
which yields naturally the relation (\ref{xamt}) between ${\mathcal T}^\mu{}_a$ and
${\mathcal J}^\mu{}_{a b}$.

On the other hand, we have shown that the minimal coupling prescription associated with the
Lorentz transformation (\ref{lt2}), that is, the coupling prescription given by a derivative
covariant under the Lorentz transformation (\ref{lt2}), yields exactly the coupling prescription
of general relativity, provided the identification
\be
A^{a}{}_{b \mu} \, x^b = c^{-2} \, B^a{}_\mu
\label{axb}
\ee
be made. This identification implies that the tetrad field and the spin connection are not
independent fields. As a consequence, the {\em local} symmetry group of general relativity
can not be the Poincar\'e group because in this group there are {\em ten} independent
parameters $\epsilon^a$ and $\epsilon^{a b}$, and ten independent gauge fields $B^a{}_\mu$ and
$A^{a}{}_{b \mu}$. The true {\em local} symmetry group behind general relativity, therefore, is
the six--parameter Lorentz group. In the form (\ref{lt3}), the Lorentz transformation of a matter
field resembles a Poincar\'e transformation, but due to the {\em four} constraints (\ref{rela}),
it is actually a transformation of the Lorentz group. In fact, if the local symmetry group were
given by the Poincar\'e group, the tetrad and the spin connection would be independent fields.

We have also seen that the tetrad field appears naturally in the theory as a consequence of the
reduction of the {\em orbital} Lorentz generator $L_{a b}$ to a translation in the coupling
prescription. The resulting tetrad field,
\be
h^a{}_\mu = \partial_\mu x^a + A^{a}{}_{b \mu} \, x^b,
\ee
is a functional of the spin connection, which reduces to the usual form (\ref{tetrada}) when the
identification (\ref{axb}) is used. In agreement with the fact that the local symmetry group of
general relativity is the Lorentz group, therefore, we can then say that the fundamental field of
gravitation is the spin connection and not the tetrad.

\acknowledgements

The authors thank R. Aldrovandi for useful discussions. They also
thank CNPq-Brazil and FAPESP-Brazil for financial support.

\end{article}

\begin{thebibliography}{99}

\bibitem[\protect\citeauthoryear{Aldrovandi and Pereira}{1995}]{livro}
Aldrovandi, R. and Pereira, J.~G. (1995).
\newblock {\em An Introduction to Geometrical Physics},
\newblock World Scientific, Singapore.

\bibitem[\protect\citeauthoryear{de Andrade and Pereira}{1997}]{trans}
de Andrade, V.~C. and Pereira, J.~G. (1997).
\newblock {\em Physical Review D} {\bf 56}, 4689.

\bibitem[\protect\citeauthoryear{Dirac}{1958}]{dirac}
Dirac, P.~A.~M. (1958).
\newblock In {\em Planck Festscrift}, W. Frank, ed., Deutscher Verlag der Wis\-senschaften,
Berlin.

\bibitem[\protect\citeauthoryear{Fock}{1929}]{fi2}
Fock, V.~A. (1929).
\newblock {\em Z. Physik} {\bf 57}, 261.

\bibitem[\protect\citeauthoryear{Fock and Ivanenko}{1929}]{fi1}
Fock, V.~A. and Ivanenko, D. (1929).
\newblock {\em Z. Physik} {\bf 54}, 798.

\bibitem[\protect\citeauthoryear{Hayashi}{1972}]{hayashi}
Hayashi, K. (1972).
\newblock {\em Lettera al Nuovo Cimento} {\bf 5}, 529.

\bibitem[\protect\citeauthoryear{Hehl et al}{1995}]{hehl}
Hehl, F.~W, McCrea, J.~D., Mielke, E.~W., and Ne'emann, Y. (1995).
\newblock {\em Physics Report} {\bf 258}, 1.

\bibitem[\protect\citeauthoryear{Kibble}{1961}]{kibble}
Kibble, T.~W.~B. (1961).
\newblock {\em Journal of Mathematical Physics} {\bf 2}, 212.

\bibitem[\protect\citeauthoryear{Konopleva and Popov}{1981}]{kopo}
Konopleva, N.~P. and Popov, V.~N. (1981).
\newblock {\em Gauge Fields},
\newblock Harwood, Chur.

\bibitem[\protect\citeauthoryear{Ramond}{1989}]{ramond}
Ramond, P. (1989).
\newblock {\em Field Theory: A Modern Primer}, 2nd edn.,
\newblock Addison-Wesley, Redwood.

\bibitem[\protect\citeauthoryear{Trautman}{1962}]{trautman}
Trautman, A. (1962).
\newblock In {\em Gravitation: An Introduction to Current Research}, L. Witten, ed., Wiley,
New York.

\bibitem[\protect\citeauthoryear{Wald}{1984}]{wald}
Wald, R.~M. (1984).
\newblock {\em General Relativity},
\newblock The Univ. of Chicago Press, Chicago.

\bibitem[\protect\citeauthoryear{Weinberg}{1972}]{weinberg}
Weinberg, S. (1972).
\newblock {\em Gravitation and Cosmology},
\newblock Wiley, New York.

\end{thebibliography}
\end{document}